\documentstyle[12pt,psfig]{article}
\begin{document}
\title 
{Multifractal Measures Characterized by the Iterative Map with
Two Control Parameters}
\author
{Kyungsik Kim \thanks{Electronic mail:
kskim@dolphin.pknu.ac.kr}, G. H. Kim and Y. S. Kong \\
{\em Department of Physics, Pukyong National University}, \\
{\em Pusan 608-737, Korea}}
%
%
%
%
\maketitle 
\abstract {A one - dimensional iterative map with two control
parameters, i.e. the Kim - Kong map, is proposed. Our purpose is to
investigate the characteristic properties of this map, and to discuss
numerically the multifractal behavior of the normalized first passage
time. Especially, based on the Monte Carlo simulation, the normalized
first passage time to arrive at the absorbing barrier after starting
from an arbitrary site is mainly obtained in the presence of both
absorption and reflection on a two - dimensional Sierpinski gasket. We
also discuss the multifractal spectra of the normalized first passage
time, and the numerical result of the Kim - Kong model presented will
be compared with that of the Sinai and logistic models.}
\section{Introduction}
\indent Recently, increasing interest has been paid to the behavior of
the disorder system in a variety of contexts in condensed matter
physics.$^{1-5}$ The stochastic process for the anomalous diffusion on
fractal structure and L$\acute{e}$vy fleights has extensively been the
subject of several literatures.$^{6-8}$ It provides a dynamic
framework for a precise connection between fractional diffusion
equations and fractal walks,$^{9,10}$ while the transport phenomena
for the motion of a test particle have largely been extended to the
reaction kinetics$^{5,11}$ and the strange kinetics.$^{12}$\\ \indent
One of the well-known problems associated mainly with the theory of
the random walk is the mean first passage time that defines the
average time arrivng at the absorbing barrier for the first time after
a walker initially starts from an arbitrary lattice point.
Particularly, this theoretical framework has extensively been applied
in the Sinai model related intimately to the random barrier with the
absorbing and reflecting barriers.  In previous work, the Sinai model
with asymmetric transition probabilities was studied for the mean and
mean square displacements dependent anomalously on time.$^{2-5}$ The
transport process for the mean first passage time has been addressed
by Noskowicz et al.$^{13}$ who obtained the upper and lower bounds
using the recursion - relation procedure.$^{14-17}$ In a one -
dimensional lattice with the periodic boundary condition, Kozak et
al.$^{18}$ have been treated with the first passage time, for the
specific case of using the logistic map as a pseudorandom number
generator.$^{19}$ On the other hand, there have recently been a few
heuristical and intuitive investigations exhibited for one -
dimensional iterative maps.  In this paper we will focus on the so -
called Kim - Kong map which takes the form of a new map with two
control parameters.  Such a map is meant to be a well - defined
dissipative map of the interval $( 0,1 )$, and it can be applied to
present the transition of a test particle from the chaotic sequences
of our map.  In particular, the Kim - Kong map proposed may be very
useful to extensively research the transport of quantum excitations,
directed polymer problems, and the electron localization in quantum
mechanics.$^{20-22}$ Interestingly, it may be expected that our
iterative map will be relevant to a nonlinear map on simple models of
spatial - temporal chaos in a nonconserving coupling map
lattice.$^{23-26}$\\ \indent Since the multifractals is directly
related to rigorously analyzed statistical property of the normalized
first passage time, the multifractal quantites have been intensively
studied in connection with the random walk problem in a one
-dimensional lattice with both absorbing and reflecting
barriers.$^{15,27}$ In general, multifractal investigations for
chaotic disorder systems have essentially discussed the transport
phenomena$^{28,29}$ and the random fractal structure with the time -
dependent random potential in a diffusive motion.$^{30-33}$ Very
recently Kim et al.$^{34}$ argued on multifractal behavior of the
normalized first passage time for the case where both absorbing and
reflecting barriers exist in a two - dimensional Sierpinski gasket.
To our knowledge, it is of fundamental importance for the
multifractals to be dealt with the transport process described by the
transition probability such as the Kim - Kong map.  In reality, the
Sinai, logistic, and Kim - Kong models studied in this paper have not
been fully explored up to now, as we will see. \\ \indent In this
study, we present the transport process of a given particle executed
on random walk with symmetric and asymmetric transition probabilities
on a two - dimensional Sierpinski gasket existing with both absorbing
and reflecting barriers.  Especially, the distribution of the
normalized first passage time arriving at absorbing barrier is
investigated primarily on the Kim - Kong model performing a random
walk by the transition probability such as the Kim - Kong map.  In
Sec.2, we discuss, in great detail, the characteristic feature of the
Kim - Kong map.  The convenient formula of the normalized first
passage time and the multifractals are also introduced.  In Sec.3, we
use the statistical value of the normalized first passage time to
relate the generalized dimension and the spectra.  The numerical
results of the Monte Carlo simulation for the generalized dimension
and the spectra are compared with those reported earlier.$^{27,34}$
Finally, we end with some results and conclusions.\\
\section{ Kim - Kong map and the multifractal feature for
the normalized first passage time }
\indent
First of all, in this section we explain in detail the characteristic properties
of the Kim - Kong map
and the multifractal feature of the normalized first passage time.
For our case the well - known quantity which defines the motion is
the transition probability of a test particle performing the symmetric or asymmetric
random walk.
For simplicity we assume that transition probability takes the form of a
Kim - Kong map. A new map that we consider is expressed in terms of
\\
\begin{equation}
x(n+1)  = \gamma \exp [-\beta (\log x(n))^2][1-\exp [-\beta (\log x(n))^2]],
\label{eq:a1}
\end{equation}
\\
where $\beta$ and $\gamma$ stand for the control parameters.
The transition probability can then be considered as
chaotic orbits of the Kim - Kong map, i.e. a sequence of pseudorandom numbers.
However, in next section we will use our map of Eq.(1) as the transition probability,
when a particle performs a random walk on a two - dimensional Sierpinski gasket.
Specifically Eq.(1) is a well - defined dissipative map with the interval $( 0,1 )$,
since it generates $x(n+1)$ from $x(n)$.
As parameter $\beta$ varies slightly higher than $0.2$
at $\gamma=3.78$, the
maximum value of $x(n+1)$
proceeds from the left to the right side of $x(n)$,
as noted in the results plotted in Fig.$1$.
It is as well obtained that the maximum value of
$x(n+1)$ exists near $x(n)=0.5$ for $ \beta=1.5$ and $\gamma=3.78$,
although it is no longer exact.
Hence we can describe higher chaotic sequences of iterated points
for $ \beta=0.2$ and $\gamma=3.78$ in Eq.(1),
while $x(n+1)$ takes on smaller chaotic values for $ \beta=2.0$ and
$\gamma=3.78$, independent of the initial value of $x(n)$.
As another characteristics, we now interpret
the bifurcation structure,
and discuss numerically the value
of two convergence ratios in a one - dimensional Kim - Kong map.
When the control parameter $\gamma$ is varied in the range  $0<\gamma<4$,
it leads to the onset of chaotic behavior, and
the bifurcation diagram of the Kim - Kong map for a period - $2^n$ point can
be shown in Fig. $2$.
It should be noted that the ${\gamma}_n$ and ${\alpha}_n$ sequences converge to the
value $\delta = \lim_{n\rightarrow \infty} {\delta}_n =4.66920 $ and
$\alpha = \lim_{n\rightarrow \infty}{\alpha}_n = 2.50290$, which are really the same values
as the Feigenbaum constants.$^{35,36}$\\
\indent
Next we will assume that a test particle performs a random walk of nearest - neighboring
transition on a one - dimensional lattice.
Then, using the well - known results of the generating function technique,
the mean first passage time$^{2,3}$ $\bar{T}$ satisfies the following equation :
\\
\begin{equation}
\bar{T} = \sum_{k=1}^{N-1} \frac{1}{p_k} + \sum_{k=0}^{N-2} \frac{1}{p_k}
\sum_{i=k+1}^{N-1} \prod_{j=k+1}^{i} \frac{q_j}{p_j},
\label{eq:b2}
\end{equation}
\\
where it means that a given particle jumps to right site with $p_j$ or left site
with $q_j$ after one step, since it starts at site $j$. The transition probabilities
$p_j$ and $q_j$ lead us to
have arbitrarily different values at the $j$ - th site when a particle jumps one step from a
given site to the nearest-neighbor site. The derivation of Eq.$(2)$ can be found in
details elsewhere.$^{14,15,19}$
There are $2^N$ realizations of the random walks possible for a chain of size N.  We can
explicitely calculate the mean first passage time of Eq.$(2)$, after enumerating
all the realizations.
Now let us consider the normalized first passage time $T_n$ defined by
\\
\begin{equation}
T_n = \frac{T_i - T_{min}}{T_{max}-T_{min}}
\label{eq:c3}
\end{equation}
\\
where $T_i$ is the first passage time required for the $i-th$ particle
to reach an absorbing barrier starting at an arbitrary site.
$T_{max}$ and $T_{min}$ are respectively the maximum and minimum values for
the first passage time arrivng at the absorbing barrier.
>From Eq.$(3)$ the normalized first passage time can be taken into account as
the rescaled real values on $(0,1)$. However, we shall be interested in
this nature employing
multifractal formula.\\
\indent
In order to exhibit the multifractal feature for the normalized first passage time,
we briefly repeat the definition of multifractals in the following.
If we divide the normalized first passage time into a set of $\epsilon$
as $\epsilon \rightarrow 0$,
then the generalized dimension in the multifractal structure$^{1,15}$ is represented as
\\
\begin{equation}
D_q = \lim_{\epsilon\rightarrow0}\frac{ln\sum_{i}n_{i}{p^q}_i}{(q-1)ln\epsilon} .
\label{eq:d4}
\end{equation}
\\
Here $n_i$ is the number of configurations
of the $i$ - th particle, and
$p_i$ is the probability for the number of configurations of the $i$ - th particle
arriving at the absorbing barrier.
Owing to Eq.(4),
the explicit relations between $D_q$ and $f_q$, and between  $\alpha_q$ and $D_q$
are respectively given by the following Legendre transform:
\\
\begin{equation}
f_q = q \frac{d}{dq}[(q-1)D_q]-(q-1)D_q
\label{eq:e5}
\end{equation}
\\
and
\begin{equation}
\alpha_q = \frac{d}{dq}[(q-1)D_q].
\label{eq:f6}
\end{equation}
\\
In our scheme, we will make use of Eqs.$(4) - (6)$ to find out
the multifractal behavior for the normalized first passage time
in a two - dimensional Sierpinski gasket, and these mathematical techniques lead us to
more general results.
\section{Result and conclusions}
\indent
For the investigations of the generalized dimension and the spectrum for the
normalized first passage time,
we take into account the two - dimensional random walk of a given particle,
where the number of sites having a stage $n$ in the $d$ - dimensional
Sierpinski gasket$^{37}$ takes on only $N_n = (d+1)(1+(d+1)^n )/2$.
Although the stage $n$ can be extended to large number in our compuetr simulation studies,
we conveniently restrict ourselves to $n=7$ stages in a two - dimensional Sierpinski gasket.
We will assume that an absorbing barrier is
at an ended site on the right side, while the reflecting
barriers are on all the boundary in Fig. 3.
After a particle starts at the initial point $i_0 =(11111233)$ at $n=7$ stages
on a two - dimensional Sierpinski gasket lattice,
this absorbs at a site $(33333333)$,where the  $n=0$ stage is located at $(1)$, $(2)$, and
$(3)$.
By using the transition probability of Eq.(1), a particle jumps
to the right site with $p_{1j} + \theta$ or the left site with $q_{1j} - \theta$,
and to the top site with $p_{2j} - \theta$ or the bottom site with $q_{2j} + \theta$
in four directions after a particle starts at a
site $j$ on a two - dimensional Sierpinski gasket. The disorder parameter $\theta$ is
the quantity for the biased value described by the direction of a given particle
on fractal lattice model, and we have
$p_{1j} + q_{1j} + p_{2j} + q_{2j} = 1$ by using the normalized condition on two -
dimensional
lattice model.
For this case the random walk of a particle is symmetric
for the disorder parameter $\theta=0$, and
will be a biased random walk that jumped to the
direction of the absorbing barrier for $0<\theta<\frac{1}{4}$.\\
\indent
For the sake of clarity we can obtain the multifractal behavior of
the normalized first passage time
from our simulation because of the apparent simplicity of our models.
Our simulations are performed for $3\times10^6$ particles and averaged over
$10^4$ configurations in the three models as
follows.
The first is the Sinai model that for Sinai disorder it has the symmetric transition
probability with $p_{1j}=q_{1j}=p_{2j}=q_{2j}=\frac{1}{4}$ and $\theta=0$, and the two
asymmetric probabilities with $\theta=0.05$ and $\theta=0.1$. The second is the logistic
model for the
transition probability represented in terms of the logistic function,
and the numerical simulation of logistic model can be treated in a similar way to
that of the Sinai model.
In this paper the logistic map is given by $x(n+1) = R x(n) [ 1 - x(n) ]$, where $R=3.9999$,
and the multifractal quantities of the normalized first passage time
we found was also refered to Ref.$[34]$, in detail.
Lastly, we concentrate on the Kim - Kong model carrying out efficiently
a random walk in the interesting case of
a Kim - Kong map for $ \beta=0.2$ and $\gamma=3.78$, where
the numerical generated sequences of the Kim - Kong map appear to
be wildly chaotic, as shown in Fig. $2$.
We find in the Kim - Kong model that the three asymmetric cutoff values are
$0.026145$, $0.147489$, and $0.492946$ with $\theta=0.05$,
while we obtain three asymmetric cutoff
values of $0.042893$, $0.147489$, and $0.379262$ with $\theta=0.1$, respectively.
Here the asymmetric cutoff value is defined as the quantity
used to  independently determine the jumping direction of a random walker. \\
\indent
>From now on we estimate numerically the generalized dimension and the spectrum
after finding the normalized first passage time from Eq.$(3)$ via the Monte Carlo
simulation.
The results of these calculations are summarized in Table I.
At present, of further interest is the fractal dimension among multifractal quantities.
Here the fractal
dimenion $D_0$ (i.e., the maximum value of $D_q$ or $f_q$ ) and the scaling exponent
$\alpha_
{\pm\infty}$, $D_{\pm\infty}$ for $\epsilon=10^{-2}$ are calculated in the Kim - Kong,
logistic, and Sinai models$^{29}$, ultimately based on the theoretical expression
Eqs.$(4)-(6)$.
The normalized first passage time in the Kim - Kong model is found to be infinite for
$\theta=0$,
while the fractal dimension in the logistic model is expected to take a value near
zero as $\theta\rightarrow0$.
It is also found that the fractal dimension
anomalously changes as the disorder parameter $\theta$ increases for
both the Kim - Kong and logistic models.
Fig. $4$ is the plot of the spectrum
$f_q$ versus $\alpha_q$  for the Sinai model on a two - dimensional Sierpinski gasket.\\
\indent
In conclusion, we have discussed in this paper the characteristic properties
of the Kim - Kong map with two control parameters.
We also have investigated the multifractals
from the distribution of the normalized first passage time arriving at
an absorbing barrier for the first time.
For completeness we have also included the results for
three models, as summarized in Table I.
In the future, the decay process
can be presented analytically and numerically from the randomness of our map
in a reaction - diffusion system with multi - species reactants.$^{38,39}$
It is expected that the detail description of
the multifractal behavior will be used to study the higher - dimensional
extensions of the fractal lattice models,
for the case of
the transition probability as a chaotic orbits of the Kim - Kong map. \\
\\
\section*{Acknowledgments}
\indent
This work is supported in part by the Academic Research Fund
of Pukyong National Univ. of Korea. \\

%
%
%
%
%
\newpage
\section*{Figures}
\begin{figure}[ht]
\centerline{\psfig{figure=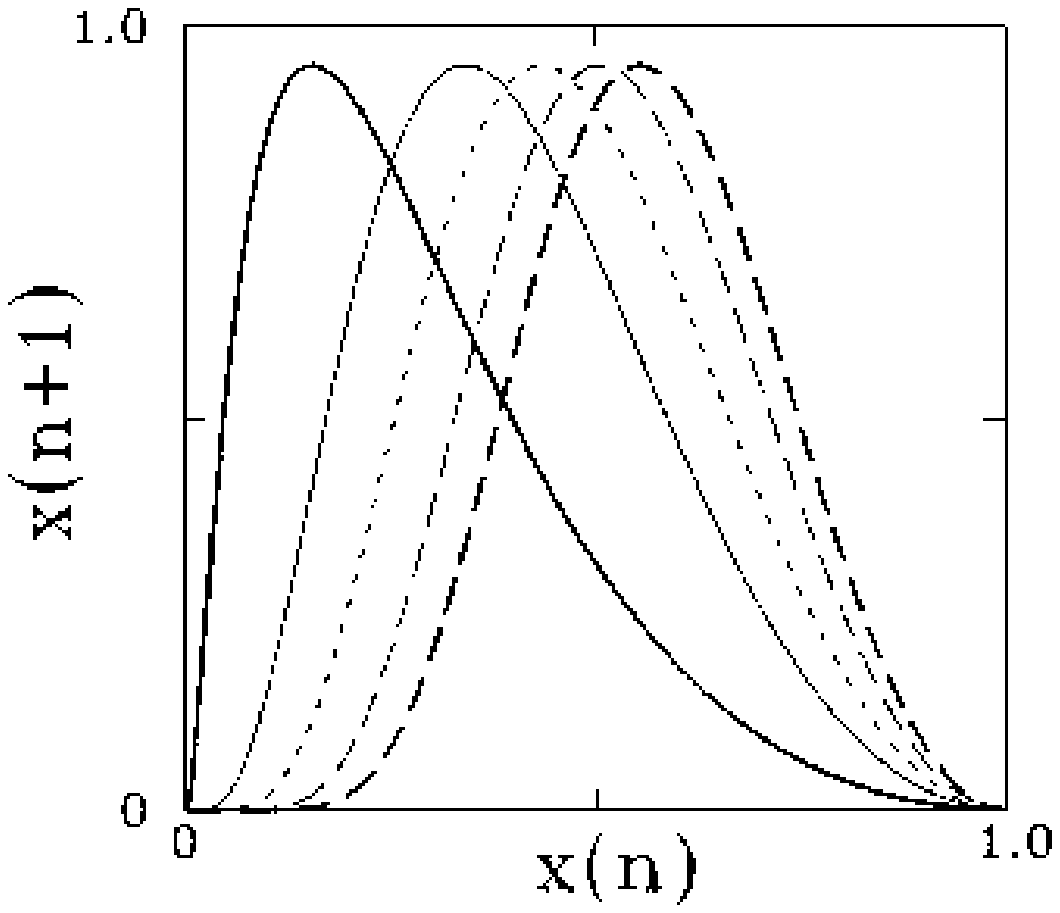,width=0.9\textwidth}}
\caption{A plot of $x(n+1)$ verse $x(n)$ graph in the Kim - Kong
map. This map is given by $x(n+1) = \gamma \exp [-\beta (\log
x(n))^2][1-\exp [-\beta (\log x(n))^2]]$, where $\beta = 0.2$, $0.6$,
$1.0$, $1.5$, and $2.0$ for $\gamma=3.78$ are respectively represented
by the thick solid, thin solid, the dot, thin dashed, and thick dashed
lines.}
\label{fig.1}
\end{figure}
\begin{figure}[hb]
\centerline{\psfig{figure=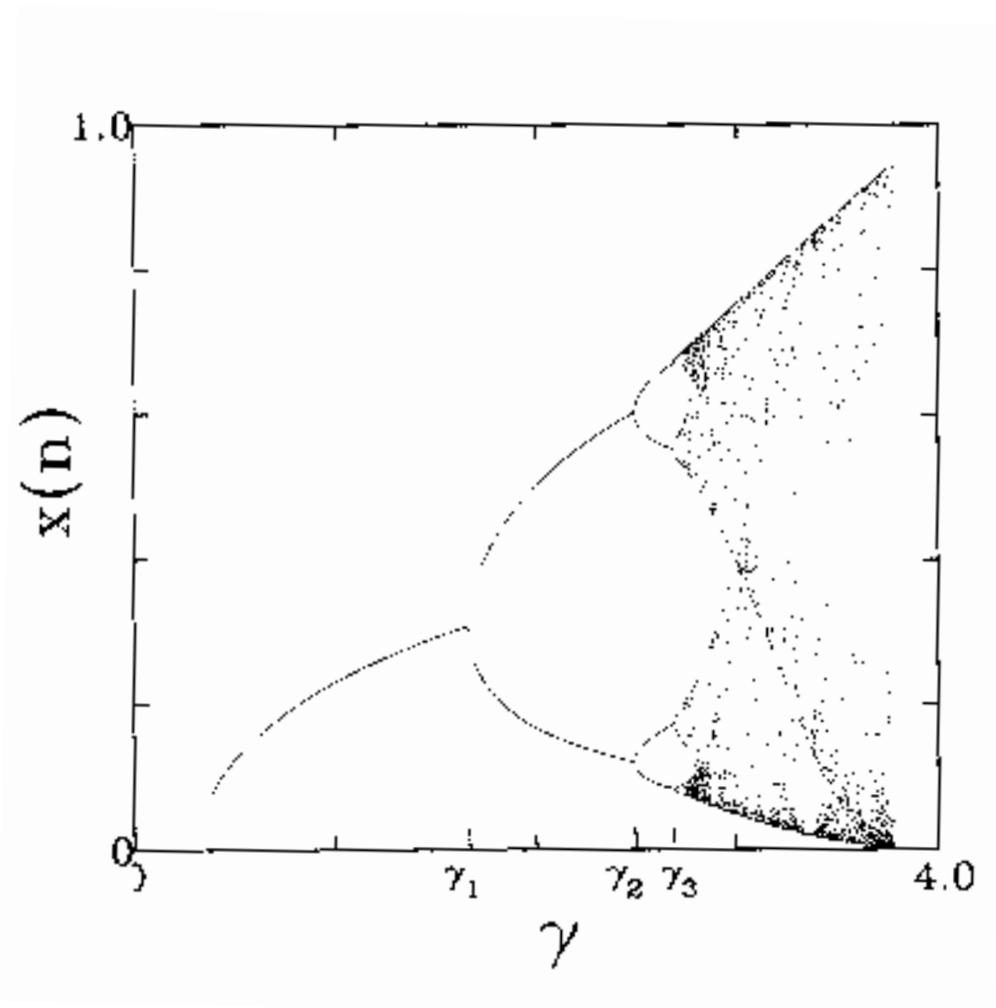,width=0.9\textwidth,angle=-1.2}}
\caption{A birfurcation diagram of the Kim - Kong map where a control
parameter $\gamma$ is varied in the range $0<\gamma<4$ at $\beta=0.2$,
and the bifurcation values ${\gamma}_1=1.673505$,
${\gamma}_2=2.502200$, and $\gamma_3=2.699329$.}
\label{fig.2}
\end{figure}
\begin{figure}[ht]
\centerline{\psfig{figure=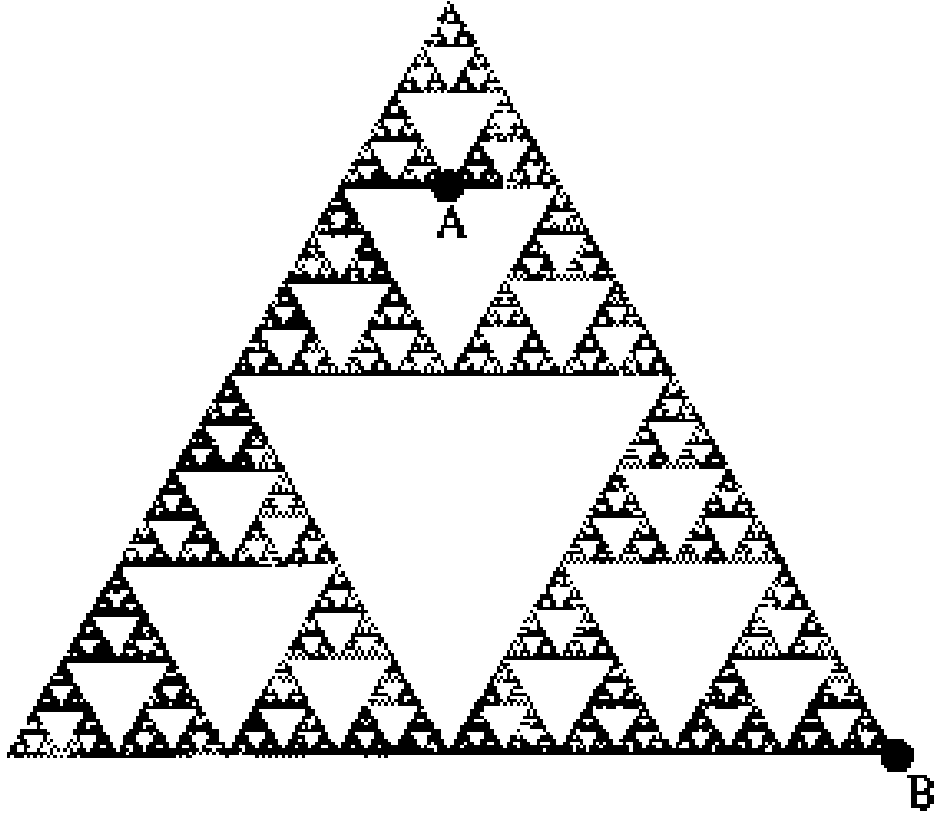,width=0.9\textwidth}}
\caption{The trajectory for a particle performing a random walk in the
logistic model with $\theta=0$, where the logistic map is given by
$x(n+1) = R x(n) [ 1 - x(n) ]$ with $R=3.9999$.  A particle is
initially started from $A=(11111233)$, and absorbed finally at
$B=(33333333)$ in a two - dimensional Sierpinski gasket at $n=7$
stages.}
\label{fig.3}
\end{figure}
\begin{figure}[ht]
\centerline{\psfig{figure=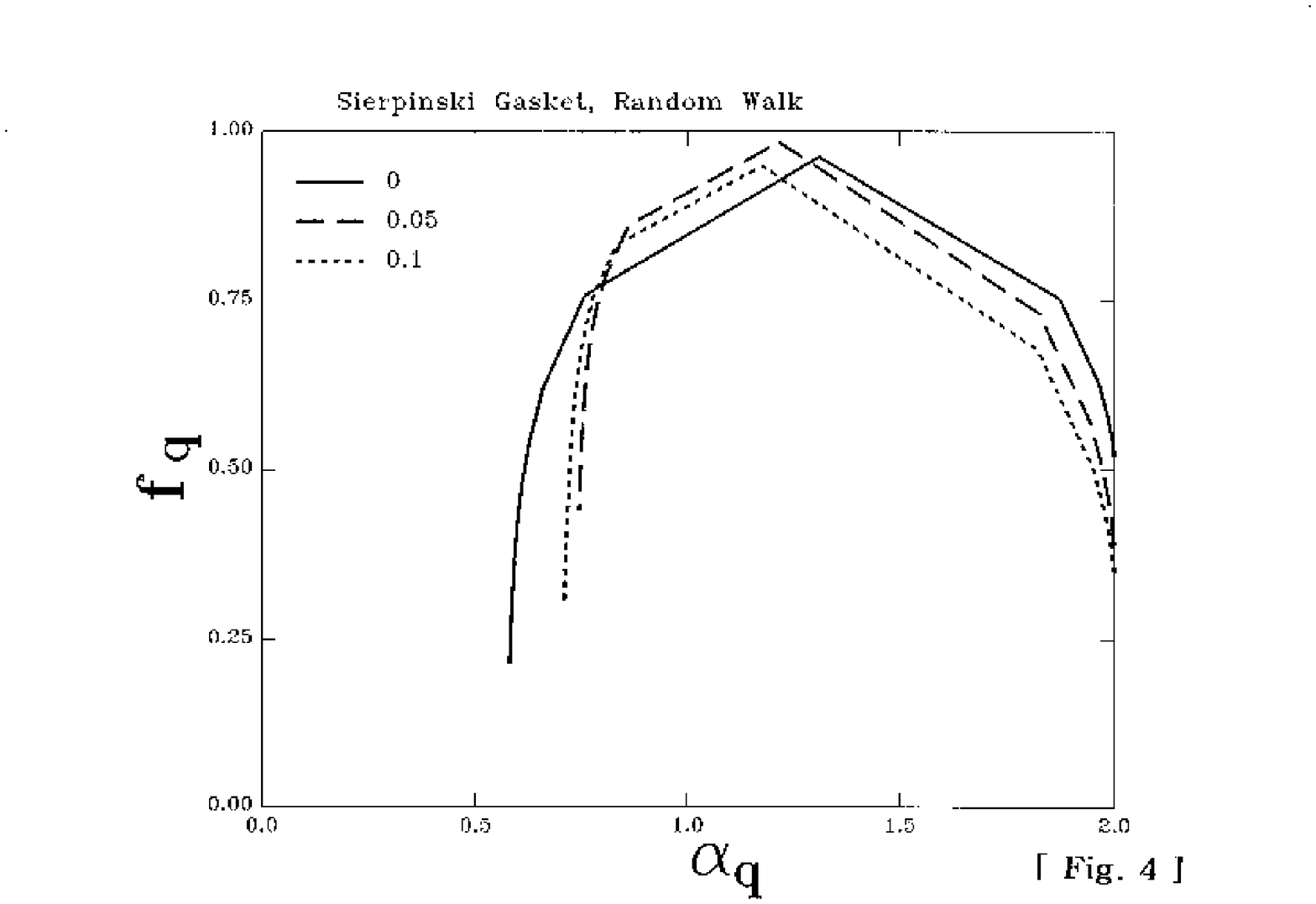,width=0.9\textwidth}}
\caption{Plots of $f_q$ and $\alpha_q$ for the disorder parameter
$\theta=0$(solid line), $0.05$(dashed line), and $0.1$(dot line)
in the Sinai model on a two - dimensional Sierpinski gasket
from Ref.$[34]$.}
\label{fig.4}
\end{figure}

\begin{figure}[ht]
\centerline{\psfig{figure=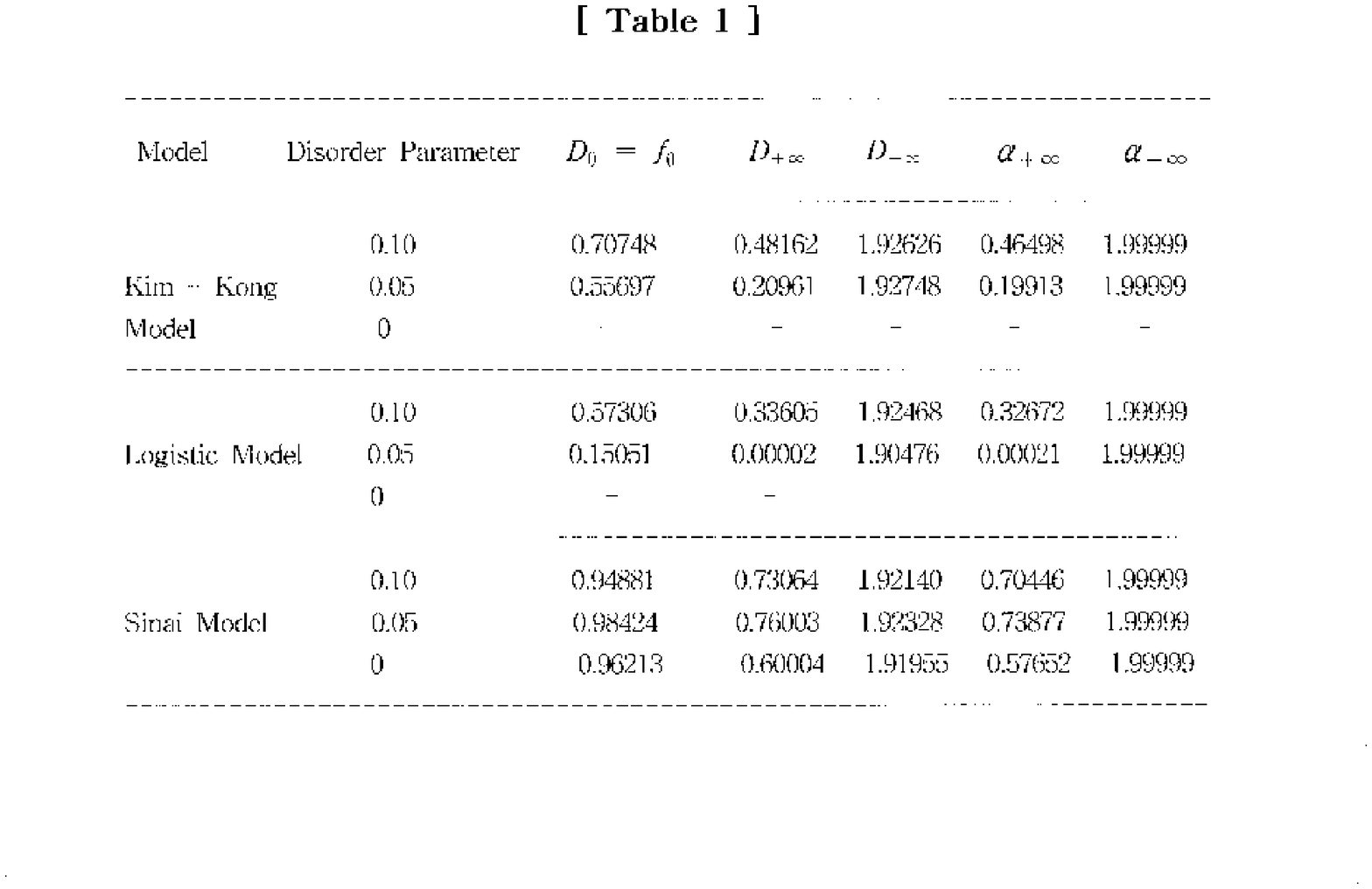,width=0.9\textwidth}} {Summary of
values of the fractal dimension , the generalized dimension, and the
spectrum in the Kim - Kong, logistic, and Sinai models on a two -
dimensional Sierpinski gasket.}
\label{table.1}
\end{figure}
\vspace{0.5cm}

\end{document}